# A WIRE POSITION MONITOR SYSTEM FOR THE 1.3 GHZ TESLA-STYLE CRYOMODULE AT THE FERMILAB NEW-MUON-LAB ACCELERATOR*

N. Eddy[#], B. Fellenz, P. Prieto, A. Semenov, D. C. Voy, M. Wendt,
Fermilab, Batavia, IL 60510, U.S.A.

*Abstract*

The first cryomodule for the beam test facility at the Fermilab New-Muon-Lab building is currently under RF commissioning. Among other diagnostics systems, the transverse position of the helium gas return pipe with the connected 1.3 GHz SRF accelerating cavities is measured along the ~15 m long module using a stretched-wire position monitoring system.

An overview of the wire position monitor system technology is given, along with preliminary results taken at the initial module cooldown, and during further testing. As the measurement system offers a high resolution, we also discuss options for use as a vibration detector.

## INTRODUCTION

An electron beam test facility, based on superconducting RF (SRF) TESLA-style cryomodules is currently under construction at the Fermilab New-Muon-Lab (NML) building [1]. The first, so-called type III+, cryomodule (CM-1), equipped with eight 1.3 GHz nine-cell accelerating cavities was recently cooled down to 2 K, and is currently under RF conditioning. The transverse alignment of the cavity string within the cryomodule is crucial for minimizing transverse kick and beam break-up effects, generated by the high-order dipole modes of misaligned accelerating structures. An optimum alignment can only be guaranteed during the assembly of the cavity string, i.e. at room temperatures. The final position of the cavities after cooldown is uncontrollable, and therefore unknown.

A wire position monitoring system (WPM) can help to understand the transverse motion of the cavities during cooldown, their final location and the long term position stability after cryo-temperatures are settled, as well as the position reproducibility for several cold-warm cycles. It also may serve as vibration sensor, as the wire acts as a high-Q resonant detector for mechanical vibrations in the low-audio frequency range [2], [3].

The WPM system consists out of a stretched-wire position detection system, provided with help of INFN-Milano [4] and DESY Hamburg, and RF generation and read-out electronics, developed at Fermilab.

## THE WIRE POSITION MONITORING SYSTEM

### The Cryomodule WPM Detection Assembly

Figure 1 gives an overview of the WPM system. It is based on a stretched wire in a coaxial transmission-line arrangement. A 28 mm inner diameter tube with bellows between seven (only four are shown in Fig. 1) stripline position pickup's, distributed along the module, serves as outer conductor, while the 0.5 mm diameter Cu-Be wire is the inner conductor of this coaxial line, giving a characteristic impedance of ~240 Ω. The wire is fixed at the feed- and endcap of the cryomodule, which are both room temperature parts of the cryo-vessel, thus reference the wire position. The wire is stretched applying the tension of an 18 kg weight over a wheel fixed to the endcap, and therefore its sag is independent of the elongation effects of the cryomodule. The WPM electronics supply a 325 MHz sine-wave signal into the stretched wire, using 1-to-5 impedance matching transformers at both ends.

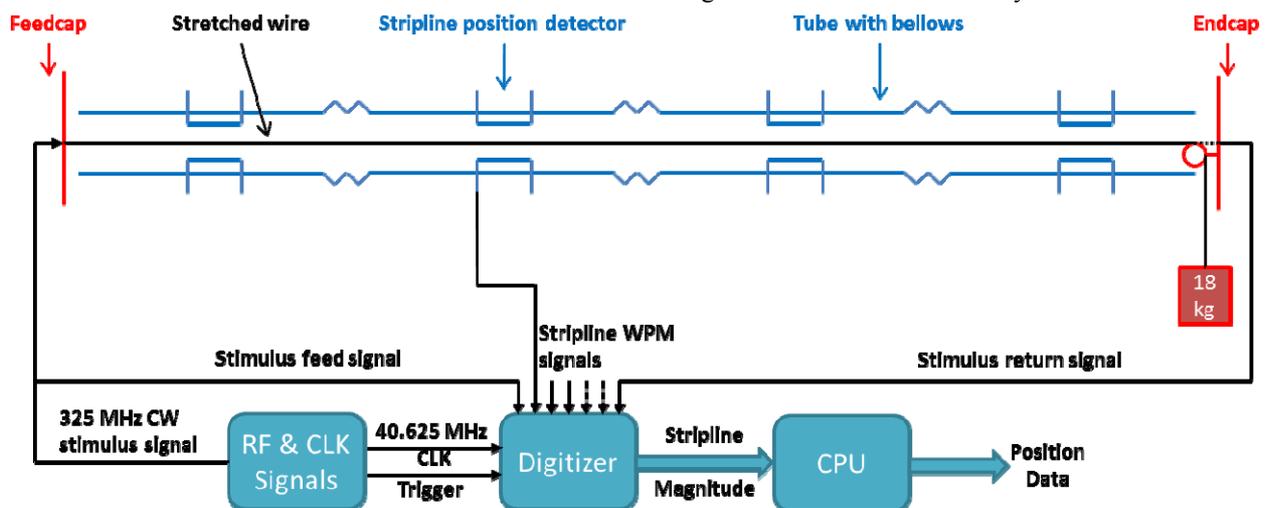

Figure 1: Schematic of the wire position monitoring system, installed at the Fermilab TESLA-style CM-1 cryomodule.

___________________________________________
* This work was supported by Fermi National Accelerator Laboratory, operated by Fermi Research Alliance, LLC under contract No. DE-AC02-07CH11359 with the United States Department of Energy
[#]eddy@fnal.gov

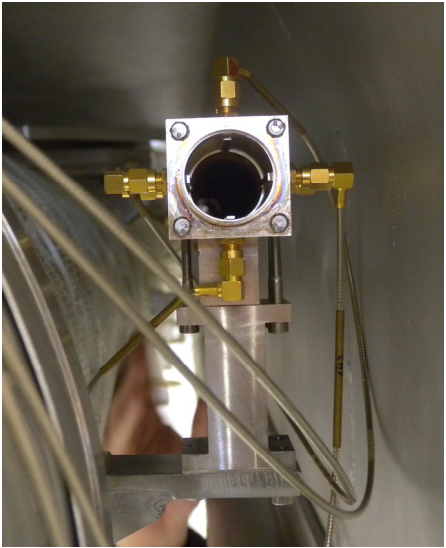

Figure 2: Stripline wire position pickup fixed to the He gas return pipe of the TESLA-style cryomodule.

The measurement of the transverse position of the wire with respect to the four symmetrically arranged electrodes of the stripline pickup (see Figure 2) is similar to a beam position monitor (BPM), i.e. the electromagnetic coupling between the wire and each electrode is a function of the transverse position of the wire. The position characteristic of the two horizontal or vertical electrodes can be approximated analytically [5], or numerically by solving the Laplace equation

$$\Delta \Phi = 0 \implies \Phi = f(x,y) \quad (1)$$

of the 2D cross-section. Figure 3 shows the equipotentials for the horizontal plane

$$\phi_h = \frac{\phi_R - \phi_L}{\phi_R + \phi_L} \quad (2)$$

keeping in mind that $\phi_h = \phi_v$ for our circular cross section.

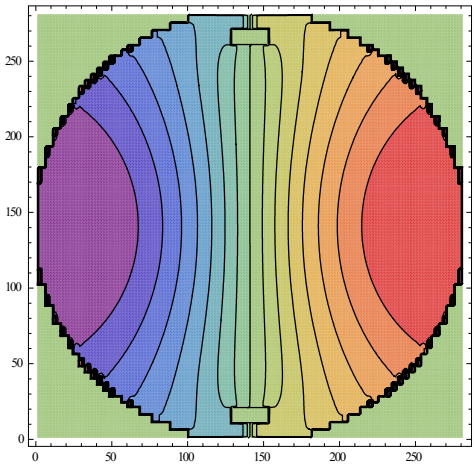

Figure 3: Horizontal position characteristic of the stripline WPM detector as of eq. (1) and (2).

A 7th-order 2D polynomial fit was applied to linearize and scale the result of eq. (2), around the pickup center the position sensitivity is ~2.6 dB/mm.

For a single cryomodule the wire is spanning a distance of 13.75 m between the reference fixtures at feed- and endcap. The tensile strength of the Cu-Be wire of ~1300 N/mm$^2$ could be further improved by a temperature annealing procedure. This allows minimizing the wire sag [6]

$$y(z) = -\frac{T}{w}\left[\cosh\left(\frac{wl}{2T}\right) - \cosh\frac{w}{T}\left(\frac{l}{2} - z\right)\right] \quad (3)$$

by applying a heavy tension weight of ~18 kg. At $z=l/2$ the sag is $y_{single} \approx$ -2.2 mm, but for a series of three cryomodules in a row ($l$ = 38 m) it increases substantially, to $y_{triple} \approx$ -16.7 mm.

*The WPM Electronics*

The transfer impedance of the WPM detector stripline electrodes computes to

$$Z_{strip}(\omega) = i Z_0 \exp\left(-i\frac{\omega l_{strip}}{c_0}\right)\sin\left(\frac{\omega l_{strip}}{c_0}\right) \quad (4)$$

with a characteristic impedance of $Z_0 \approx 50~\Omega$, and a length l = 50 mm, the frequency range of the stripline electrode is $f_{3dB} \approx$ 750…4500 MHz. At the selected operating frequency $f$ = 325 MHz the amplitude level is ~9.5 dB below the maximum achievable level at $f_{center}$ = 1500 MHz. However, this compromise was made to simplify the RF electronics without using an analog down-converter.

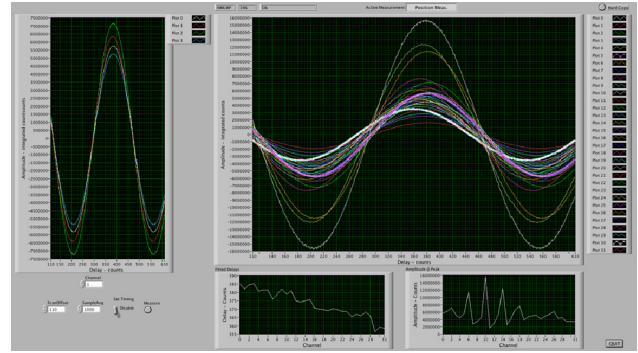

Figure 4: Delay scan results for all 32 digitizer channels. One delay step is 9 ps.

As Figure 1 indicates, a digitzer is the central component of the WPM electronics, here in form an in-house developed 8-ch 125 MSPS 14-bit VME board with FPGA based signal processing. The AD6445 ADC-chip provides 500 MHz analog bandwidth, which allows an undersampling of the 325 MHz CW input signal at a synchronous sampling frequency, e.g. 81.25 MHz, 40.625 MHz, etc. A PECL-based precision delay circuit for each digitizer, as part of the VME RF & clock generation board, allows the remote controlled adjustment of the sampling clock to the peak values of the 325 MHz sine-wave input signal. A timing scan is performed every 5 minutes to account for slow phase drifts. The sampling clock to each 8 channel digitizer can be adjusted 9 ps steps. A typical delay scan is shown in Figure 4. In this way we rectify the RF input signal at the ADC, however, the control of latencies and individual cable delay differences proved to be critical as they determine the delay spread within each 8 channel digitizer. We also included a 24 dB RF gain stage for each stripline electrode (located in the tunnel), in order to match the low level output signal of the pickup to the 1 $V_{pp}$ full scale range of the digitizer.

Before commissioning of the complete WPM system at the CM-1 cryomodule, some experience was gained at a test setup [7]. This also allowed optimizing the decimation and filter parameters in the FPGA-based digital signal processing [8]. Table 1 lists the parameters of the current setup. After filtering and decimation, the magnitude data of each electrode is forwarded to the VME Motorola 5500 CPU, which runs the WPM data acquisition software under VxWorks. Here the linearization and scaling is performed according to the 2D polynomial fit of eq. (2), as well as the control of all VME digitizers and the RF & clock module.

Table 1: Parameters of the WPM signal processing

| CW input frequency $f_{in}$ | 1300 MHz / 4 = 325 MHz |
|---|---|
| ADC clock frequency $f_{CLK}$ | $f_{in}$ / 8 = 40.625 MHz |
| Average & decimate filter | N = 9918 samples |
| Effective sample frequency $f_{SR}$ | $f_{CLK}$ / N = 4.096 kHz |
| # of samples per $t_{trig} \approx 4$ sec | 16384 |
| Frequency resolution | ~0.25 Hz |

## PRELIMINARY RESULTS

*Slow Motion*

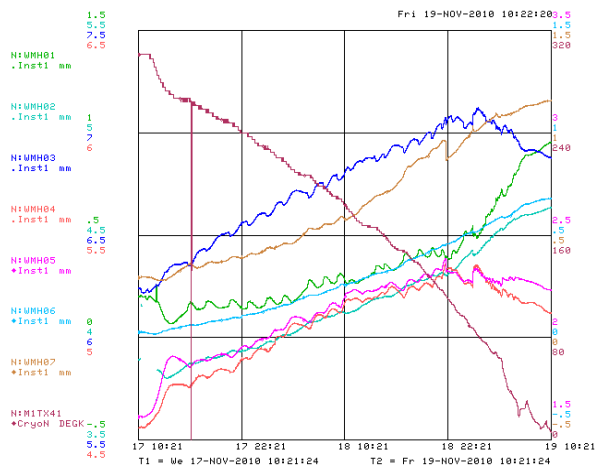

Figure 5: Horizontal motion of the pickups during initial 48 hour cooldown from room temperature to 4 K.

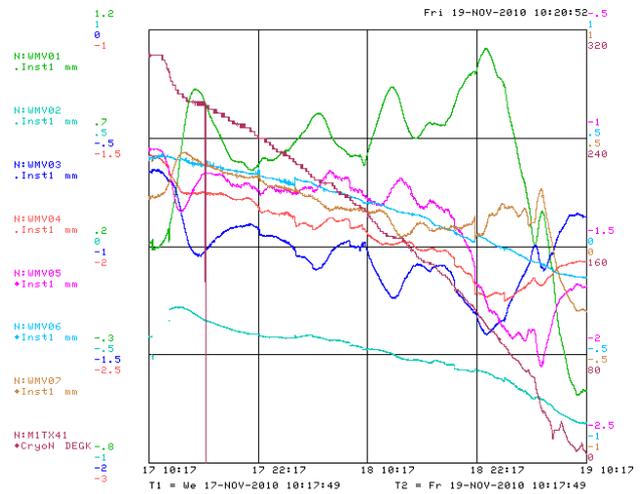

Figure 6: Vertical motion of the pickups during initial 48 hour cooldown from room temperature to 4 K.

The WPM system provides feedback on the slow motion of the cavities in the module, which is the primary purpose of this measurement system. During the very first initial cooldown of the CM-1 cryomodule, performed at Fermilab, the WPM system received a lot of attention while monitoring the behaviour of the module during the cooldown process. The position at each pickup is data logged by the accelerator controls system every 15 seconds. The observed motion at each pickup over the inital 48 hour cooldown from room temperature to 4 K is shown in Fig. 4 and Fig. 5, for horizontal and vertical respectively. For horizontal, there is general positive drift to the module with some swaying as it cools down. For vertical, the return pipe undergoes contortions about the fixed supports approximately located at pickups 2, 4, and 6. The largest motion is observed at the first pickup near the feedcap end of the pipe. Note, this end of the pipe is attached via a bellows and not rigidly fixed to the endcap. While cold, the module has remained quite stable. Fig.6 and Fig. 7 show the horizontal and vertical motion of the module as the temperature was cycled between 2 to 4 K.

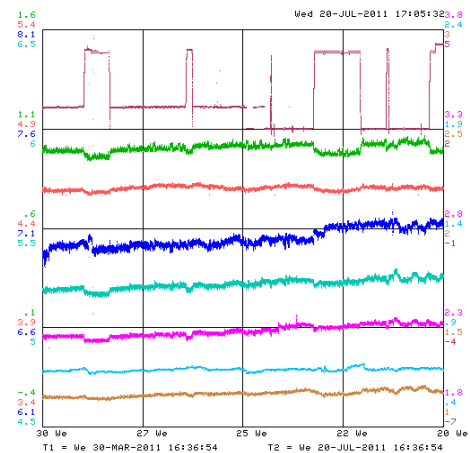

Figure 7: Horizontal motion of the pickups over 4 months as the temperature cycled from 2 to 4 K.

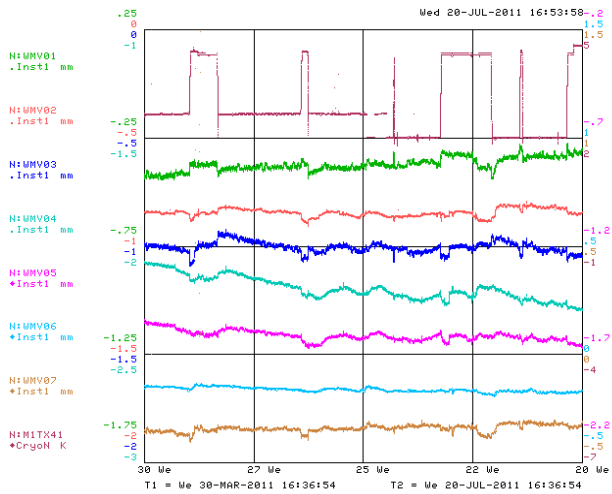

Figure 8: Vertical motion of the pickups over 4 months as the temperature cycled from 2 to 4 K.

*Fast Motion - Vibration*

The WPM system can also provide fast motion feedback with a sampling frequency of 4 kHz and up to 16K samples. This provides vibration data up to 2 kHz with a frequency resolution of 0.25 Hz. Any mechanical excitation will cause the wire to vibrate as given by the vibrating wire equation

$$f_n = \frac{n}{2l}\sqrt{\frac{F}{\rho A}} \quad (5)$$

Using eq. (5), with a wire of length $l$ = 13.75 m, density $\rho$ = 8.26x10$^3$ kg/m$^3$, and radius $r$ = 0.254 mm, under a tension of 18.144 kg, yields a fundamental frequency of 11.75 Hz. A preliminary check of the wire vibration shows that each wire is indeed vibrating at harmonics of 11.75 Hz as shown in Fig. 8.

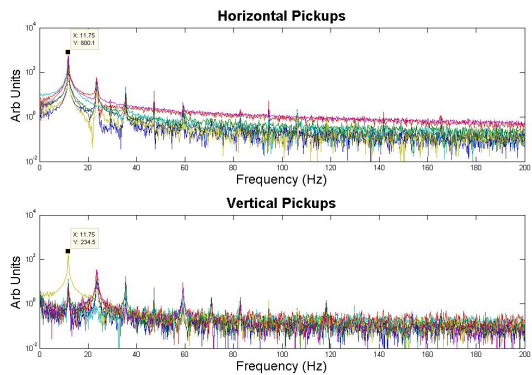

Figure 9: Preliminary vibration analysis. The first 200Hz span of the spectrum is shown.